\documentstyle[12pt]{article}

\newlength{\dinwidth}
\newlength{\dinmargin}
\newcommand{\ba}{\begin{array}}
\newcommand{\ea}{\end{array}}
\newcommand{\be}{\begin{equation}}
\newcommand{\ee}{\end{equation}}
\newcommand{\bea}{\begin{eqnarray}}
\newcommand{\eea}{\end{eqnarray}}
\newcommand{\beas}{\begin{eqnarray*}}
\newcommand{\eeas}{\end{eqnarray*}}

\def\g{\gamma}

\def\to{\rightarrow}
\def\bee{\begin{eqnarray}}
\def\eee{\end{eqnarray}}
\def\be{\begin{equation}}
\def\ee{\end{equation}}

\def\laplace{{\kern1pt\vbox{\hrule height 1.2pt\hbox{\vrule width
1.2pt\hskip
  3pt\vbox{\vskip 6pt}\hskip 3pt\vrule width 0.6pt}\hrule height 0.6pt}
  \kern1pt}}
\def\scriptlap{{\kern1pt\vbox{\hrule height 0.8pt\hbox{\vrule width
0.8pt
  \hskip2pt\vbox{\vskip 4pt}\hskip 2pt\vrule width 0.4pt}\hrule height
0.4pt}
  \kern1pt}}

\def\roughly#1{\raise.3ex\hbox{$#1$\kern-.75em\lower1ex\hbox{$\sim$}}}

\def\g2ym{\bar{g}^2_{\scriptscriptstyle YM}}

\begin{document}
\thispagestyle{empty}
\addtocounter{page}{-1}
\begin{flushright}
CLNS 97/1490\\
IASSNS-HEP 97-60\\
SNUTP 97-073\\
{\tt hep-ph/9708355}\\
\end{flushright}
\vspace*{1.3cm}
\centerline{\Large \bf Chiral Perturbation Theory for Tensor Mesons
\footnote{
Work supported in part by NSF Grant, NSF-KOSEF
Bilateral Grant, KOSEF Purpose-Oriented Research Grant 94-1400-04-01-3
and SRC-Program, Ministry of Education Grant BSRI 97-2410, the Monell
Foundation and the Seoam Foundation Fellowships.}}
\vspace*{1.2cm}
\centerline{\large\bf Chi-Keung Chow${}^a$ and Soo-Jong Rey${}^{b,c}$}
\vspace*{0.6cm}
\centerline{\large\it Newman Laboratory for Nuclear Studies}
\vskip0.1cm
\centerline{\large\it Cornell University, Ithaca NY 14853 USA${}^a$}
\vskip0.3cm
\centerline{\large\it School of Natural Sciences, Institute for Advanced
Study}
\vskip0.1cm
\centerline{\large\it Olden Lane, Princeton NJ 08540 USA${}^b$}
\vskip0.3cm
\centerline{\large\it Physics Department, Seoul National University,
Seoul 151-742 KOREA${}^c$}
\vspace*{0.4cm}
\centerline{\large\tt ckchow@lnsth.lns.cornell.edu, sjrey@ias.edu}
\vspace*{1.5cm}
\centerline{\large\bf abstract}
\vskip0.5cm
Interactions of $a_2, K^*_2, f_2$ and $f_2'$ tensor-mesons with
low-energy $\pi, K, \eta, \eta'$ pseudo-scalar mesons are constrained
by chiral symmetry. We derive a chiral Lagrangian of tensor mesons
in which the tensor mesons are treated as heavy non-relativistic matter 
fields. Using $1/N_c$ counting, we derive relations among unknown
couplings 
of the chiral Lagrangian. 
Chiral perturbation theory is applied to the tensor-meson mass matrix. 
At one-loop there are large corrections to the individual tensor
meson masses, but the singlet-octet mixing angle remains almost 
unchanged. We argue that all heavy mesons of spin $\ge 1$ share common
feature of chiral dynamics. 
\vspace*{1.1cm}

\centerline{\tt Submitted to Physics Letters B}

\setlength{\baselineskip}{18pt}
\setlength{\parskip}{12pt}
\newpage

An important application of chiral perturbation theory is to describe
the 
interaction of matter fields (such as nucleons \cite{1,2} or hadrons 
containing a heavy quark \cite{3,4,5,6}) with low-momentum
pseudo-Goldstone 
bosons -- the pions, kaons and eta.  
In Ref.~\cite{7} chiral perturbation theory was used to describe the 
interactions of the $\rho$, $K^*$, $\phi$ and $\omega$ vector mesons
with 
low-momentum pseudo-Goldstone bosons.  
In this article, we will extend the formalism to study the lowest-lying 
tensor meson nonet.  
This nonet is expected to have quantum numbers $J^P = 2^+$, and contains 
the isotriplet $a_2(1320)$ and the $S=\pm 1$ isodoublets $K_2^*(1430)$.  
The two isosinglet states are not as well established, but experimental 
evidences suggest that they are probably $f_2(1275)$ and $f'_2(1525)$.  
The mass difference between these nine lowest-lying
tensor mesons are small compared to the chiral symmetry breaking scale 
$ 4 \pi f_\pi \approx 1$ GeV. Hence, chiral perturbation theory should be 
applicable as a systematic expansion procedure for a class of 
processes involving tensor mesons and soft Goldstone bosons. In the
past, chiral perturbation theory has been used extensively to study 
processes which do not have a tensor meson in the final state.  
In such decays, the final state pions are not soft enough that the 
application of the chiral Lagrangian to such processes is not justified
a priori. At the best, they serve as a phenomenological model. 

The pseudo-Goldstone boson fields can be written as a $3\times3$ special 
unitary matrix 
\begin{equation}
\Sigma = \exp {2i\Pi\over f}, 
\end{equation}
where 
\begin{equation}
\Pi = \left(\matrix{
{\pi^0\over\sqrt{2}}+{\eta\over\sqrt{6}}&\pi^+&K^+\cr
\pi^-&-{\pi^0\over\sqrt{2}}+{\eta\over\sqrt{6}}&K^0\cr
K^-&\overline K^0&-{2\eta\over\sqrt{6}}\cr}\right).
\end{equation}
Under chiral $SU(3)_L \times SU(3)_R$, $\Sigma\to L\Sigma R^\dagger$,
where 
$L\in SU(3)_L$ and $R\in SU(3)_R$.  
At leading order in chiral perturbation theory, $f$ can be identified
with 
the pion or kaon decay constant ($f_\pi \sim 132$ MeV, $f_K \sim 160$
MeV).  
It is convenient, when describing the interactions of the
pseudo-Goldstone 
bosons with other fields to introduce 
\begin{equation}
\xi = \exp {i\Pi\over f} = \sqrt{\Sigma}.   
\end{equation}
Under chiral $SU(3)_L \times SU(3)_R$, 
\begin{equation}
\xi\to L\xi U^\dagger = U\xi R^\dagger, 
\end{equation}
where in general $U$ is a complicated function of $L$, $R$ and the meson 
fields $\Pi$.  
For transformations $V=L=R$ in the unbroken $SU(3)_V$ subgroup, $U=V$.  

The tensor meson fields are introduced as a $3\times 3$ octet matrix 
\begin{equation}
\tilde {\cal O}_{\mu\nu} = \left(\matrix{
{a_2^0\over\sqrt{2}}+{f_2^{(8)}\over\sqrt{6}}&a_2^+&K_2^{*+}\cr
a_2^-&-{a_2^0\over\sqrt{2}}+{f_2^{(8)}\over\sqrt{6}}&K_2^{*0}\cr
K_2^{*-}&\overline
K_2^{*0}&-{2f_2^{(8)}\over\sqrt{6}}\cr}\right)_{\mu\nu}, 
\end{equation}
and as a singlet 
\begin{equation}
\tilde {\cal S}_{\mu\nu} = {f_2^{(0)}}_{\mu\nu}.  
\end{equation}
(We have deliberately made our notations for the tensor mesons
identical 
to that for the vector mesons in Ref.~\cite{7} except for an additional 
tilde.)
By definition these tensor mesons are symmetric and traceless Lorentz 
tensors, 
\begin{eqnarray}
\tilde {\cal O}_{\mu\nu} &= \tilde {\cal O}_{\nu\mu},
\qquad \tilde {\cal O}^\mu_\mu &= 0,\nonumber\\
\tilde {\cal S}_{\mu\nu} &= \tilde {\cal S}_{\nu\mu},
\qquad \tilde {\cal S}^\mu_\mu &= 0.  
\end{eqnarray}
Moreover, the polarizations of the tensor mesons are necessarily
orthogonal 
to the momentum, 
\begin{equation}
p^\mu \tilde {\cal O}_{\mu\nu} = p^\mu \tilde {\cal S}_{\mu\nu} = 0.  
\label{orth}
\end{equation}
Under chiral $SU(3)_L\times SU(3)_R$, 
\begin{equation}
\tilde {\cal O}_{\mu\nu} \to U \tilde {\cal O}_{\mu\nu} U^\dagger, 
\qquad \tilde {\cal S}_{\mu\nu} \to \tilde {\cal S}_{\mu\nu},  
\end{equation}
and under charge conjugation, 
\begin{equation}
C\tilde {\cal O}_{\mu\nu}C^{-1} = \tilde {\cal O}_{\mu\nu}^T, \qquad
C\tilde {\cal S}_{\mu\nu}C^{-1} = \tilde {\cal S}_{\mu\nu}, \qquad
C\xi C^{-1} = \xi^{T}.  
\end{equation}

We construct a chiral lagrangian for tensor mesons by treating the
tensor mesons as heavy static fields \cite{8,9} with fixed four-velocity
$v_\mu$, 
with $v^2 = 1$.  
Eq.~(\ref{orth}) becomes 
\begin{equation}
v^\mu \tilde {\cal O}_{\mu\nu} = v^\mu \tilde {\cal S}_{\mu\nu} = 0.  
\end{equation}
The chiral lagrangian density which describes the interactions of the 
tensor mesons with the low-momentum $\pi$, $K$ and $\eta$ mesons has the 
general structure 
\begin{equation}
{\cal L} = {\cal L}_{\rm kin} + {\cal L}_{\rm int} + {\cal L}_{\rm
mass}.  
\end{equation}
At the leading order in the derivative and quark mass expansions, 
\begin{equation}
{\cal L}_{\rm kin} = -{i \over 2}
\tilde{\cal S}_{\mu\nu}^\dagger (v\cdot \partial) 
\tilde{\cal S}^{\mu\nu} - {i \over 2} 
\,{\rm Tr}\, \tilde{\cal O}_{\mu\nu}^\dagger (v\cdot {\cal D}) 
\tilde{\cal O}^{\mu\nu}, 
\end{equation}
and 
\begin{eqnarray}
{\cal L}_{\rm int} &=& \textstyle{i \over 2}
\tilde g_1 \tilde{\cal S}_{\mu\rho}^\dagger \,{\rm Tr} \,
(\tilde{\cal O}_\nu^\rho A_\lambda) v_\sigma
\epsilon^{\mu\nu\lambda\sigma} 
+ h.c. \nonumber\\ &&\; 
+ \textstyle{i \over 2} \, \tilde g_2 \,{\rm Tr}\, 
(\{\tilde{\cal O}_{\mu\rho}^\dagger, \tilde{\cal O}_\nu^\rho\}
A_\lambda) 
v_\sigma \epsilon^{\mu\nu\lambda\sigma}, 
\end{eqnarray}
where 
\begin{equation}
{\cal D}_\lambda \tilde{\cal O}_{\mu\nu} = \partial_\lambda 
\tilde{\cal O}_{\mu\nu} + [V_\lambda, \tilde{\cal O}_{\mu\nu}], 
\end{equation}
and 
\begin{equation}
V_\lambda = \textstyle{1\over2}(\xi \partial_\lambda \xi^\dagger + 
\xi^\dagger \partial_\lambda \xi), \qquad 
A_\lambda = \textstyle{i\over2}(\xi \partial_\lambda \xi^\dagger - 
\xi^\dagger \partial_\lambda \xi).  
\end{equation}
Finally, to linear order in quark mass expansion, 
\begin{eqnarray}
{\cal L}_{\rm mass}=
&& {\tilde\mu_0 + \tilde\sigma_0 \,{\rm Tr} \, {\cal M} \over 2} \, 
\tilde{\cal S}_{\mu\nu}^\dagger \tilde{\cal S}^{\mu\nu} 
+ {\tilde\mu_8 + \tilde\sigma_8 \,{\rm Tr} \, {\cal M} \over 2} 
\,{\rm Tr}\, \tilde{\cal O}_{\mu\nu}^\dagger \tilde{\cal O}^{\mu\nu} 
\nonumber\\ &&\quad 
+ {\tilde\lambda_1 \over 2}
\,{\rm Tr}\, (\tilde{\cal O}_{\mu\nu}^\dagger 
{\cal M}_\xi) \tilde{\cal S}^{\mu\nu} + h.c.
+ {\tilde\lambda_2 \over 2}
\,{\rm Tr}\, (\{\tilde{\cal O}_{\mu\nu}^\dagger, 
\tilde{\cal O}^{\mu\nu}\} {\cal M}_\xi), 
\end{eqnarray}
where $\cal M$ is the quark mass matrix ${\cal M} =$ diag$(m_u, m_d,
m_s)$, 
and 
\begin{equation}
{\cal M}_\xi = \textstyle{1\over2} (\xi{\cal M}\xi
+ \xi^\dagger{\cal M}\xi^\dagger).
\end{equation}
As mentioned in Ref.~\cite{7}, one of the mass parameters can be removed
by 
a simultaneous phase redefinition of $\cal O$ and $\cal S$, and only the 
singlet-octet mass difference $\Delta\tilde\mu\equiv\tilde\mu_0 -
\tilde\mu_8$ 
is physically relevant.  
It is yet unclear which $f_2$ state corresponds to the lowest-lying
isosinglet 
tensor meson, but for all reasonable choices $\Delta\tilde\mu \leq 300$
MeV 
and can be regarded as a quantity of order $m_q$.  
Note that $\Delta\tilde\mu$ is of order $N_c^{-1}$ and vanishes in the
large 
$N_c$ limit.

To fix couplings in the chiral Lagrangian, we analyze the spectrum
of tensor mesons given at leading order in chiral perturbation theory.
The analysis is essentially identical to the SU(3) analysis. In the
approximation with exact isospin symmetry $m_u = m_d = \hat m$, we 
find that $a_2$ and $K^*_2$ tensor mesons are interaction and energy
eigenstates simultaneously and have masses as: 
\bee
M_{a_2} &=& {\overline \mu}_8 + 2 {\tilde \lambda}_2 \hat m
\nonumber \\
M_{K^*_2} &=& {\overline \mu}_8 + {\tilde \lambda}_2 (\hat m + m_s)
\eee
while the $f^{(0)}$ and $f^{(8)}$ mesons have mass matrix
\be
M^{(8-0)} 
= \left( \begin{array}{cc}
{\overline \mu}_8 + {2 \over 3} {\tilde \lambda}_2 (\hat m + 2 m_s)
& {2 \over \sqrt 6} {\tilde \lambda}_1 (\hat m - m_s) \\
{2 \over \sqrt 6} {\tilde \lambda}_1^* (\hat m - m_s) &
{\overline \mu}_0 \end{array} \right).
\ee 
We have abbreviated combinations of parameters as
\be
{\overline \mu}_8 = \tilde\mu_8 + \tilde\sigma_8 {\rm Tr}\, {\cal M}
\hskip1cm 
{\overline \mu}_0 = \tilde\mu_0 + \tilde\sigma_0 {\rm Tr} \,{\cal M}.
\ee
Using the above relations, we find that the singlet-octet mass matrix
elements can be expressed in terms of the experimentally measured 
tensor-meson masses:
\bee
M_{11}^{(8-0)}
&=& {4 \over 3} M_{K^*_2} - {1 \over 3} M_{a_2} 
\nonumber \\
M_{22}^{(8-0)} &=&
M_{f_2'} + M_{f_2} - {4 \over 3} M_{K^*_2} + {1 \over 3} M_{a_2}
\nonumber \\
M_{12}^{(8-0)} = M_{21}^{(8-0)}
&=& \pm \Big[
\Big( {4 \over3 } M_{K^*_2} - {1 \over 3} M_{a_2} - M_{f_2'} \Big) 
\cdot \Big( M_{f_2} - {4 \over 3} M_{K^*_2} + {1 \over 3} M_{a_2}
\Big) \Big].
\eee
The mass eigenstates of octet-singlet mixing tensor mesons are 
parametrized by a mixing angle $\Theta_T$:
\be
\left( \begin{array}{c} |f_2 \rangle \\ |f_2' \rangle \end{array}
\right)
= 
\left( \begin{array}{cc} 
\cos \Theta_T & + \sin \Theta_T \\
- \sin \Theta_T & \cos \Theta_T \end{array} \right)
\left( \begin{array}{c} |f_2^{(8)} \rangle \\
|f_2^{(0)} \rangle \end{array} \right).
\ee
The above relation of octet-singlet mass matrix suggests the usual
SU${}_V$(3) prediction for the tangent of the mixing angle
\be
\tan \Theta_T = \mp \sqrt{ {M_{f_2'} - {4 \over 3} M_{K^*_2} 
+ {1 \over 3} M_{a_2} \over {4 \over 3} M_{K^*_2} - {1 \over 3} M_{a_2} 
- M_{f_2} }} = \mp 0.556 \approx \mp {1 \over \sqrt{3}}.
\label{mix}
\ee
Here, we have identified $M_{f_2} = 1270 $ MeV and $M_{f_2'} = 1525$
MeV respectively. 
This translates into a mixing angle $\Theta_T = 29.1^\circ$, which is 
close but not exactly the value $\theta_{\rm lin} = 26^\circ$ as quoted 
in Particle Data Group \cite{10}.  
For comparison $\Theta_T = 35^\circ$ for an ideal mixing, and
experimentally 
$\Theta_T$ is measured to be $(25.3 \pm 1.1)^\circ$ \cite{11}.  

In the large $N_c$ limit, quark loops are suppressed. Thus the leading
diagrams in the meson sector contain a single quark loop. As a result,
the octet and the singlet tensor mesons can be combined into a single
{\sl nonet} matrix
\be
\tilde{\cal N}_{\mu \nu} = \Big( \tilde{\cal O} + { {\bf 1} \over \sqrt
3} 
\tilde{\cal S} \Big)_{\mu \nu}.
\ee
The chiral Lagrangian in the large $N_c$ limit is expressed exclusively
in terms of $\tilde{\cal N}_{\mu \nu}$ tensor field. At leading order in 
$1/N_c$, the kinetic and interaction parts are given by
\bee
{\cal L}_{\rm kin} &=& - \textstyle{i \over 2}
\, {\rm Tr} \tilde{\cal N}^\dagger_{\mu \nu}
\, v \cdot {\cal D} \tilde{\cal N}^{\mu \nu}
\nonumber \\
{\cal L}_{\rm int} &=& {i {\tilde g}_2 \over 2} \, {\rm Tr} 
\big( \{ \tilde{\cal N}^\dagger_{\mu \rho}, \tilde{\cal N}^\rho_\nu \}
A_\lambda \big) v_\sigma \epsilon^{\mu \nu \lambda \sigma}
\eee
while the mass matrix part is given by
\be
{\cal L}_{\rm mass} = {\tilde\mu \over 2}
{\rm Tr} \tilde{\cal N}^\dagger_{\mu \nu} \tilde{\cal N}^{\mu \nu}
+ { {\tilde \lambda}_2 \over 2}
\, {\rm Tr} \Big( \{ \tilde{\cal N}^\dagger_{\mu \nu},
\tilde{\cal N}^{\mu \nu} \} {\cal M}_\xi \Big).
\ee
Comparison with the chiral Lagrangian we have constructed above
indicates that in the $N_c \rightarrow \infty$ limit
\be
\Delta {\tilde \mu} \rightarrow 0, \hskip0.5cm {\tilde \sigma}_0 
\rightarrow 0, \hskip0.5cm {\tilde \sigma}_8 \rightarrow 0,
\ee
\be
{\tilde g}_1 \rightarrow {2 {\tilde g}_2 \over \sqrt{3}}, 
\hskip0.5cm 
{\tilde \lambda}_1 \rightarrow {2 \lambda_2 \over \sqrt{3}}
\hskip0.5cm 
\tan \Theta_T \rightarrow {1 \over \sqrt{2}}.
\ee
This means that $|f_2 \rangle$ state becomes `pure' $(s {\overline s})$
state and the nonet matrix is given by
\be
{\cal N}_{\mu \nu} = \left( \begin{array}{ccc}
{a_2^0 \over \sqrt 2} + {f_2 \over \sqrt 2} & a_2^+ & K_2^{*+} \\
a_2^- & - {a_2^0 \over \sqrt 2} + {f_2 \over \sqrt 2} & K_2^{*0} \\
K_2^{*-} & \overline{K}^{*0}_2 & f_2' \end{array} \right).
\ee

In leading order one has the ideal mixing angle 
$( \Theta_T )_{N \rightarrow \infty} = 35^\circ$ which is
close 
but off by 20\% from the actual value $\Theta_T = 29.1^\circ$ obtained 
from Eq.~(\ref{mix}).  
Note, however, that the mixing angle is quite sensitive to the $f_2$
masses, 
and Eq.~(\ref{mix}) can be brought into consistency with ideal mixing by 
just shifting $f'_2(1525)$ up by 40 MeV.  
This discrepancy may be due to contamination of  $f_2 - f_2'$ tensor
mesons 
with exotic states such as tensor glueball states and/or exotic 
mesons such as $\rho-\rho$, 
$\omega-\omega$ bound-state resonances \cite{12}. 

The coupling constants $\tilde g_{1,2}$ are free parameters in 
chiral perturbation theory.  
To estimate these coupling constants, we decompose the quark-model
wave functions of vector and tensor mesons in their rest frames 
($v=(1, \vec 0)$) into their isospin, spin and spatial parts.  
\begin{eqnarray}
|V^a(\epsilon^i)\rangle &=& |\lambda^a\rangle 
\otimes |S=1, S_z = i\rangle \otimes |1s, L=0, L_z = 0\rangle, 
\nonumber\\
|T^a( \epsilon^{ij} \rangle &=& |\lambda^a \rangle 
\otimes |S=1, S_z = i\rangle \otimes |2p, L=1, L_z = j\rangle 
+ (i\leftrightarrow j) - \hbox {\rm trace in $(ij)$}.    
\end{eqnarray}
Note that the vector and tensor mesons are different only in 
their spatial wave functions.  
Current algebra \cite{13, 14} suggests that the axial current 
coupling is given by 
\begin{equation}
A^{ai} \sim V \lambda^a {\bf S}^i V^\dagger, 
\end{equation}
where 
\begin{equation}
V = \exp(-i\theta ({\bf S} \times {\bf L})_z), 
\end{equation}
with $\theta$ a small parameter ($\sin^2 \theta \sim {1\over8}$ in the
baryon 
sector, and is expected to be of similar magnitude for mesons).  
In the zeroth order of an expansion in $\theta$, the axial current 
coupling $A^{ai} = \lambda^a S^i$ does not couple to the spatial 
wave functions at all.  
As a result, $g_{1,2} = \tilde g_{1,2}$, {\it i.e.}, the same 
parameters govern the axial couplings of the vector and tensor 
mesons.  
As discussed in \cite{7}, under the assumption of ideal mixing, 
$\tilde g_1 = g_1 = 2/\sqrt{3}$ and $\tilde g_2 = g_2 = 1$ in the 
nonrelativistic constituent quark model.  
If we employ the non-relativistic chiral quark model~\cite{15}, these 
factors are further reduced by a factor of $3/4$.

In chiral perturbation theory, the leading order corrections to the 
tensor meson masses are of order $m_q^{3/2}$ and arise from one-loop
self-energy diagrams due to virtual pseudoscalar meson exchange. A
straightforward calculation gives
\bee
\delta M_{a_2} &=& - {1 \over 8 \pi f^2} \Big[{\tilde g}_2^2
\Big( {2 \over 3} m_\pi^3 +  2 m_K^3 + {2 \over 3} m_\eta^3 \Big) 
+ {\tilde g}_1^2 m_\pi^3 \Big]
\nonumber \\
\delta M_{K^*_2} &=& - {1 \over 8 \pi f^2} \Big[ {\tilde g}_2^2 
\Big({3 \over 2} m_\pi^3 + {5 \over 3} m_K^3 + {1 \over 6} m_\eta^3
\Big) + {\tilde g}_1^2 m_K^3 \Big]
\nonumber \\
\delta M^{(08)}_{11} 
&=& - {1 \over 8 \pi f^2} {\tilde g}_1^2 \Big( 3 m_\pi^3 + 4 m_K^3 
+ m_\eta^3 \Big)
\nonumber \\
\delta M^{(08)}_{22} 
&=& - {1 \over 8 \pi f^2} \Big[ {\tilde g}_2^2 \Big( 2 m_\pi^3
+ {2 \over 3} m_K^3 + {2 \over 3} m_\eta^3 \Big) + {\tilde g}_1^2 
m_\eta^3 \Big]
\nonumber \\
\delta M^{(08)}_{12} &=& \delta M^{(08)}_{21} 
= + {1 \over 8 \pi f^2} \sqrt{2 \over 3}
{\tilde g}_1 {\tilde g}_2
( - 3 m_\pi^3 + 2 m_K^3 + m_\eta^3).
\eee
The mass corrections are quite substantial. For example, $\delta
m_{a_2} \approx - 450$ MeV. 
On the other hand, the singlet-octet mixing angle $\Theta_T$ 
after leading order chiral perturbation one-loop correction is taken
into account remains very small. The mixing angle is
\be
\tan \Theta_T = \mp \sqrt{
M_{f_2'} - {4 \over 3} M_{K_2^*} + {1 \over 3} M_{a_2} - \delta M
\over 
{4 \over 3} M_{K_2^*} - {1 \over 3} M_{a_2} - M_{f_2} + \delta M}, 
\label{ang}
\ee
where 
\bee
\delta M &=& - {4 \over 3} \delta M_{K_2^*} + {1 \over 3} \delta M_{a_2}
+ \delta M_{22}^{(08)}
\nonumber \\
&=& - { 1 \over 8 \pi f^2} 
\Big({\tilde g}_1^2 + {2 \over 3} {\tilde g}_2^2 \Big) 
\Big( {1 \over 3} m_\pi^3 - {4 \over 3} m_K^3 + m_\eta^3 \Big).
\eee
Using the large $N_c$ relation between 
${\tilde g}_1$ and ${\tilde g}_2$, we find that  
\be
\delta M
 \approx - {2 {\tilde g}_2^2 \over 8 \pi f^2}
\Big( {1 \over 3} m_\pi^3 - {4 \over 3} m_K^3 + m_\eta^3 \Big).
\label{mass}
\ee
For a reasonable range ${\tilde g}_2 \approx 0.75 - 1$, the mass 
correction $\delta M \approx - 6$ MeV.  
The corresponding shift in mixing angle is about $1^\circ$. 
Its smallness (when contrasted with the huge corrections to individual 
masses) is presumably because the combination $\delta M$ has to vanish
in 
the large $N_c$ limit, {\it i.e.}, $\delta M = {\cal O}(N_c^{-1})$. 

A remark is in order. Various heuristic arguments indicate that $J=2$
tensor mesons might exhibit moderate mixing with other $2^{++}$ meson
states. The first is tensor glueball. In the past, there has been
various
phenomenological model for the mixing. It has been argued that, once
the glueball mixing is taken into account, the $f_2$ meson is in ideal
mixing \cite{11}. 
The second are tetra-quark states. There has been no systematic
analysis of their influence to the mixing. Within chiral perturbation 
theory the tetraquark states might be treatable systematically as a 
perturbation of $({\bf 8}_L, {\bf 8}_R) \subset ({\bf 8}_L , {\bf 8}_R) 
\otimes ({\bf 8}_L, {\bf 8}_R)$ irreducible state. 
Electromagnetic correction to the mass matrix is of theoretical
interest.
In this case one has to bear in mind that both short-distance and 
long-distance effects contributions have to be taken into account.

It is straightforward to generalize the chiral perturbation theory
to $(q {\overline q})$ mesons of higher spin. The spin-$s$ meson fields
are introduced in terms of $3 \times 3$ octet matrix ${\cal O}_{\mu_1 
\mu_2 \cdots \mu_s}$ and a singlet ${\cal S}_{\mu_1 \mu_2 \cdots 
\mu_s}$. The spin-$s$ mesons are irreducible representations of the
Lorentz group, viz. totally symmetric and traceless components. The 
polarization of these mesons are necessarily orthogonal to the momentum. 
For heavy static fields the four-velocity also satisfies 
$v^{\mu_1} {\cal O}_{\mu_1 \mu_2 \cdots \mu_s} = 
v^{\mu_1} {\cal S}_{\mu_1 \cdots \mu_s} = 0$.
Chiral Lagrangian of heavy spin-$s$ tensor mesons is exactly the same
as Eqs.(13)-(18):
\bee
{\cal L}_{\rm kin} &=& - {i \over s!}
{\cal S}^\dagger_{\mu_1 \cdots \mu_s} (v \cdot \partial) {\cal S}^{\mu_1
\cdots \mu_s}
- {i \over s!} {\rm Tr} \, {\cal O}_{\mu_1 \cdots \mu_s}^\dagger
(v \cdot {\cal D}) {\cal O}^{\mu_1 \cdots \mu_s}
\nonumber \\
&&
\nonumber \\
{\cal L}_{\rm int} &=&\cases{0, &$s=0$; \cr \cr \noalign{\vskip 2pt}
{i \over s!} g_1 {\cal S}^\dagger_{\mu \rho_1 \cdots}
{\rm Tr} \, ({\cal O}^{\rho_1 \cdots}_\nu A_\lambda) v_\sigma 
\epsilon^{\mu \nu\lambda \sigma} + (h.c.) &\cr \cr
\qquad+ { i \over s!} g_2 {\rm Tr} (\{ {\cal O}_{\mu \rho_1
\cdots}^\dagger, {\cal O}^{\rho_1 \cdots}_\nu \} A_\lambda)
v_\sigma \epsilon^{\mu \nu \rho \sigma}, &otherwise,\cr}
\nonumber \\ 
& & \nonumber \\
{\cal L}_{\rm mass}
&=& 
{ {\overline \mu}_0 \over s!} {\cal S}^\dagger_{\mu_1 \cdots
\mu_s} {\cal S}^{\mu_1 \cdots \mu_s} 
+ {{\overline \mu}_8 \over s!} {\rm Tr} 
{\cal O}^\dagger_{\mu_1 \cdots \mu_s} {\cal O}^{\mu_1 \cdots \mu_s}
\nonumber \\
&& \nonumber \\
&& + {\lambda_1 \over s!} {\rm Tr} ({\cal O}^\dagger_{\mu_1 
\cdots \mu_s} {\cal M}_\xi) {\cal S}^{\mu_1 \cdots \mu_s} + (h.c.)
+ {\lambda_2 \over s!} {\rm Tr} (\{ {\cal O}^\dagger_{\mu_1 \cdots
\mu_s}, 
{\cal O}^{\mu_1 \cdots \mu_s} \} {\cal M}_\xi).
\eee  

Since the interaction Lagrangian structure is exactly the same as
$2^{++}$ tensor mesons, the leading order and the one-loop correction
to mass spectra and mixing angles follow exactly the same pattern. 
Hence the singlet-octet mixing for these higher spin mesons are also 
related to their masses through Eq.~(\ref{mix}), and be ideally mixed  
in the large $N_c$ limit.  
For example, by plugging in the masses of the spin-3 mesons one obtains 
the mixing angle $\Theta = 28^\circ$, in qualitative agreement with the 
large $N_c$ prediction.  
Moreover, the chiral one-loop correction will take the form of 
Eq.~(\ref{ang}) where $\delta M$ will be given by Eq.~(\ref{mass}) up 
of spin-dependent multiplicative constants.  
Note that, however, these higher spin mesons are heavier in mass and 
there are many other possible sources of contamination, which cannot be 
treated within the framework of chiral perturbation theory.  

In this letter, we have studied chiral perturbation theory
of heavy tensor mesons of spin $\ge 2$. We have shown that the octet
- singlet mixing angle is quite close to `ideal mixing' and one-loop
correction to the mixing angle is negligible. On the other hand, 
mass spectra themselves receive quite a sizable corrections.
We conclude that the chiral perturbation theory `explains' naturally
why the singlet-octet mixing is signficantly off the ideal mixing
for pseudo-scalar Goldstone bosons while close to the ideal mixing
for all higher spin $\ge 1$ mesons. 


We thank S.~von~Dombrowski, F. Wilczek, M.B. Wise and T.M. Yan for
useful 
discussions.

\end{document}